\documentclass[12pt]{article}
\usepackage{engl_u}
\usepackage{iheprepe_u}
\usepackage{ipsizen}
\usepackage{authors_u}

\begin{document}

\begin{titlepage}

\prepnum{2017-2}{}

\author{\small O.P. Yushchenko, V.F. Kurshetsov, A.P. Filin, S.A. Akimenko, 
A.V. Artamonov, A.M. Blik, V.V. Brekhovskikh, V.S. Burtovoy, S.V. Donskov, A.V. Inyakin, 
A.M. Gorin, G.V. Khaustov, S.A. Kholodenko, 
V.N. Kolosov, A.S. Konstantinov, V.M. Leontiev, V.A. Lishin,  M.V. Medynsky, 
Yu.V. Mikhailov,   V.F. Obraztsov, V.A. Polyakov, A.V. Popov, V.I. Romanovsky, 
 V.I. Rykalin,  A.S. Sadovsky, V.D. Samoilenko,  V.K. Semenov,  
 O.V. Stenyakin,  O.G. Tchikilev, V.A. Uvarov (NRC "Kurchatov Institute"-IHEP,
 Protvino), \\
 V.A. Duk, S.N. Filippov, E.V. Guschin,Yu.G. Kudenko, A.A. Khudyakov, V.I. Kravtsov, 
 A.Yu.Polyarush  (INR-RAS, Moscow), \\
 V.N. Bychkov, G.D. Kekelidze, V.M. Lysan,  B.Zh. Zalikhanov (JINR, Dubna)}

\title{\boldmath $K_{e3}$ decay studies in OKA experiment}

\submitted{JETP Letters}
\end{titlepage}

\begin{abstractpage}[519.25.256]
\numpag{5}
\numref{11}
\numtab{1}
\numfig{3}

\engabs{Yushchenko O.P et al.}{$K_{e3}$ decay studies in OKA experiment}
Recent results from OKA setup concerning form factor studies in $K_{e3}$ decay are presented.
 About 5.25M events are selected for the analysis. The  linear and quadratic slopes for
 the decay formfactor $f_{+}(t)$  are measured:
 $\lambda'_{+}=(26.1 \pm 0.35 \pm 0.28 )\times 10^{-3}$, $\lambda''_{+}=(1.91 \pm 0.19 \pm 0.14)\times 10^{-3}$. The
 scalar and tensor contributions are compatible with zero.
Several alternative parametrizations are tried: the Pole fit parameter is found to be 
$M_V = 891 \pm 2.0$ MeV ;  the parameter
of the Dispersive parametrization is measured to be   $\Lambda_+ =(24.58 \pm 0.18) \times 10^{-3}$.
  The presented results are considered as preliminary.

\rusabs{Ющенко О.П. и др.}{Исследования $K_{e3}$ распада в эксперименте ОКА}
Представлены новые результаты исследования $K_{e3}$ распада, осуществленные на установке ОКА.
В анализе использованы около $5.2M$ событий.
Измерянные линейный и квадритичный параметры наклона формфактора  $f_{+}(t)$: $\lambda'_{+}=(26.1 \pm 0.35 \pm 0.28 )\times 10^{-3}$, $\lambda''_{+}=(1.91 \pm 0.19 \pm 0.14)\times 10^{-3}$.
 Вклады скалярного и тензорного членов сравнимы с нулем. Использовались несколько альтернативных параметризаций: параметр полюсного фита $M_V = 891 \pm 2.0$ MeV; параметр дисперсионной параметризации
$\Lambda_+ =(24.58 \pm 0.18) \times 10^{-3}$.
 Представленные результаты являются предварительными.

\selectlanguage{english}

\end{abstractpage}

\section*{Introduction}
The kaon decays  provide  unique information about the dynamics of the strong interactions.
It has been a testing ground for such theories as current algebra, PCAC, Chiral Perturbation Theory (ChPT) etc. Another 
direction is a search for new interactions, such as tensor and scalar ones. Here, we present a high-statistics
study of $K_{e3}$ decays from OKA detector at U-70 Proton Synchrotron.

\section{OKA beam and detector}
OKA is the abbreviation for $'$Experiments with Kaons $'$. OKA beam is a RF-separated secondary beam of U-70 Proton Synchrotron 
of IHEP, Protvino. The beam is described elsewhere \cite{okabeam}. RF-separation with Panofsky scheme is realised. It uses two
superconductive  Karsruhe-CERN SC RF deflectors \cite{Sepa}, donated by CERN. Sophisticated cryogenic system, built at IHEP \cite{cryo}
provides superfluid He for cavities cooling. The resulting beam has up to $\sim 20 \% $ of kaons with an  intensity of
$\sim 10^{6}$ kaons per 3 sec U-70 spill. 
\begin{figure*}[h]
 \center {\includegraphics[width=\linewidth]{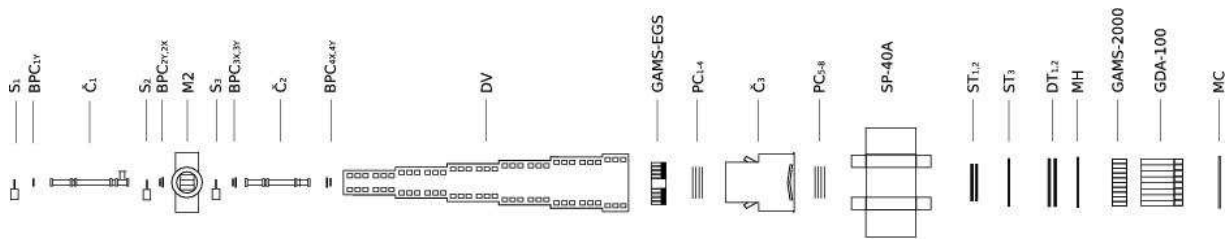}}
\caption{OKA setup} 
\label{Fig:okafigure}
\end{figure*}
The OKA setup is a magnetic spectrometer, presented on Fig.~\ref{Fig:okafigure}.
It includes:
 \begin{enumerate}
\item Beam spectrometer on the basis of 7 1mm pitch PC's ($BPC_{x,y}$) $\sim$1500 channels in total,
4 2mm-thick scintillation counters and 2 threshold Cherenkov counters.
\item Decay volume(DV) with Veto system, 11m long, filled with He, veto system is composed of
 670 Lead-Scintillator sandwiches 20$\times$(5 mm Sc + 1.5 mm Pb) with WLS readout. The counters
 are grouped in 300 ADC channels.
\item Main magnetic spectrometer: $200 \times 140~$cm$^2$ aperture magnet wih $\int{ B dl} \sim$ 1 Tm; 5K 2 mm
 pitch PC's; 1K 9 mm Straw's and 300 channels of 40 mm DT's.
\item Gamma detectors: GAMS-2000 ($\sim 2300 ~3.8\times 3.8\times 45.$ cm$^3$ lead glass blocks), large angle
 detector (EGS) ($\sim 1050 ~ 5\times 5\times 42$ cm$^3$ lead glass blocks).
\item  Muon detector: GDA-100 Hadron Calorimeter (100 $20 \times 20$ cm$^2$ iron-scintillator sandwiches with 
WLS plates readout); 4 $1 \times 1$ m$^2$ Sc counters behind GDA-100.
\end{enumerate} 
  \section{Trigger and statistics}
\label{trstat}
Very simple trigger, which is almost $''$minimum bias $''$ one,   has been used during data-taking:\\ 
$Tr= S_{1} \cdot S_{2} \cdot S_{3} \cdot \overline{ \check C}_{1} \cdot \check C_{2} 
\cdot \overline{S}_{bk} \cdot (\Sigma_{GAMS}>MIP)  $.
It is a combination of beam Sc counters,
 $\check{C}_{1,2}$ threshold Cerenkov counters ($\check{C}_{1}$ sees pions, $\check{C}_{2}$- pions and 
kaons), ${S}_{bk}$ - a $''$beam-killer $''$ counter located in the beam-hole of the GAMS 
gamma-detector.  $\Sigma_{GAMS}>MIP$ is a requirement for the analog sum of amplitudes in the 
GAMS-2000 to be higher than a MIP signal. \\
The $''$OKA$''$ is taking data since 2010, the total  available statistics  corresponds to $\sim 15M K_{e3}$
decays. In the present study we use part of the statistics taken in 2012 and 2013.
\section{$K_{e3}$ decay study.}
\label{Ke3selection}
\begin{figure}
 \begin{minipage}{0.35\linewidth}
 \center {\includegraphics[width=\linewidth]{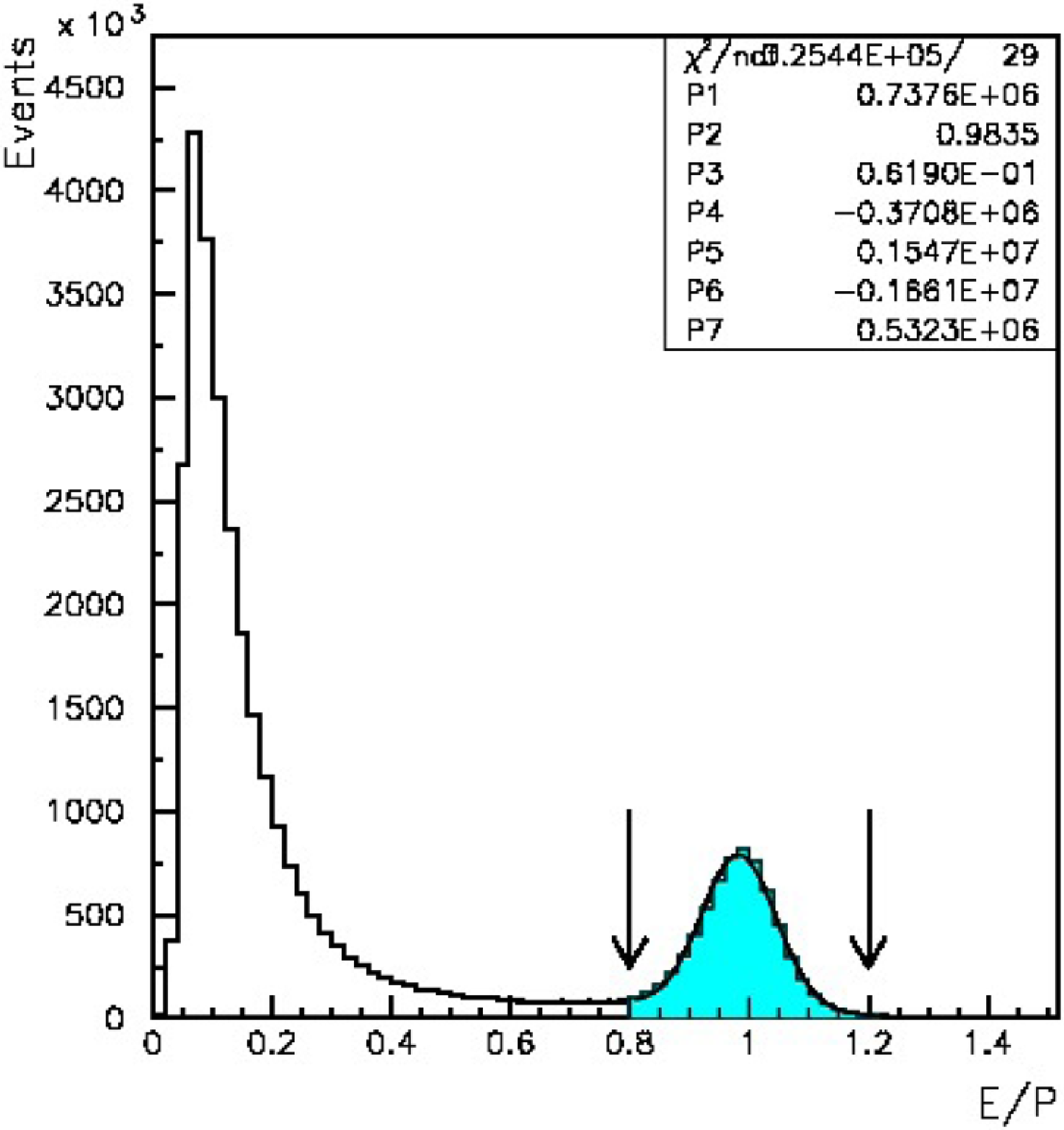}}
 \end{minipage}\hspace{2pc}%
 \begin{minipage}{0.49\linewidth}
 \center {\includegraphics[width=1.2\linewidth]{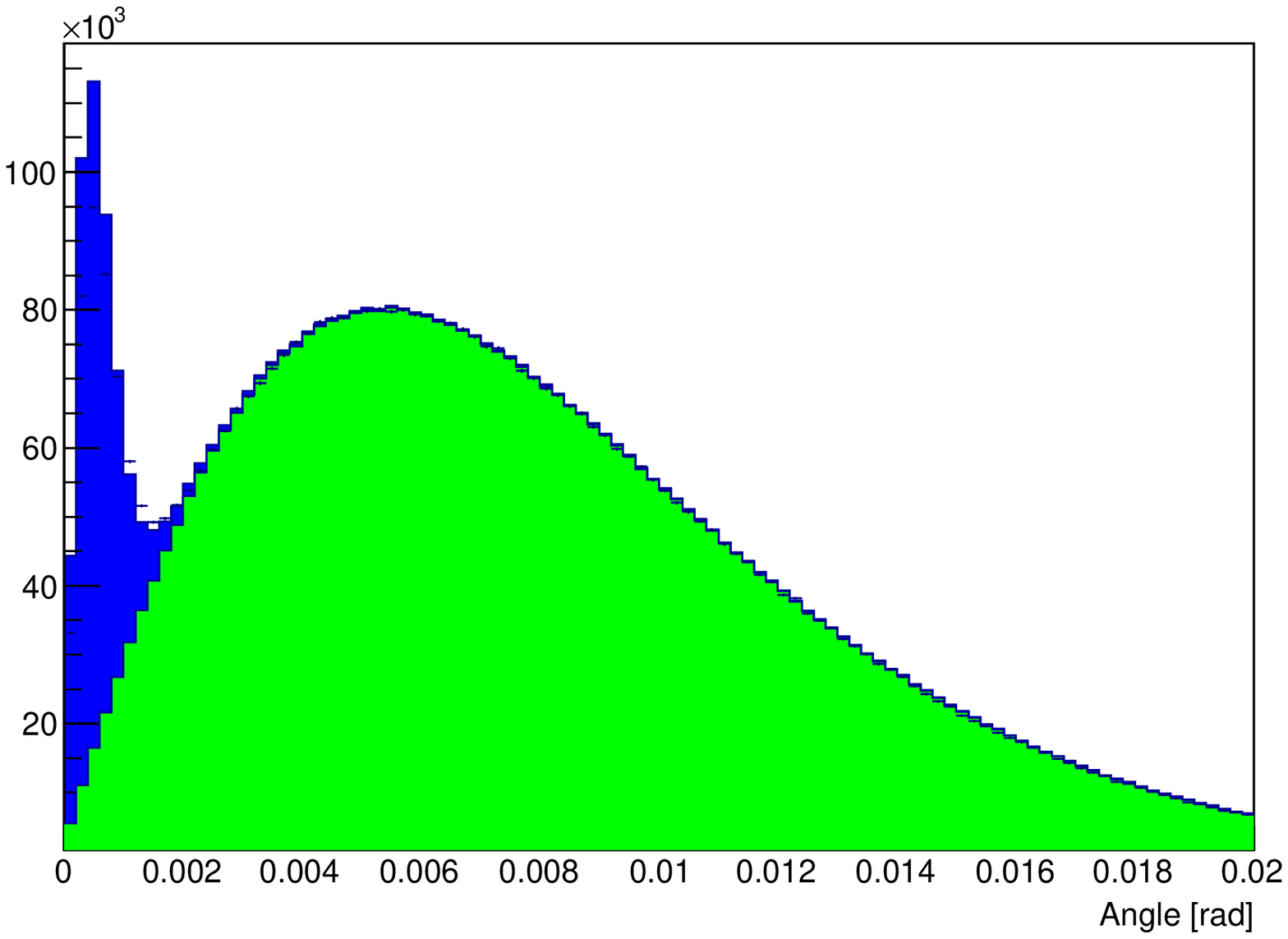}}\vspace{0.05pc}%
 \end{minipage}
\caption{E/P plot - the ratio of the energy of the associated ECAL
 cluster to the momentum of the charged track (left); $\alpha$ - the angle between $\vec p_{K}$ and $\vec p_{e}+ \vec p_{\pi}$ 
in the lab-system (right).} 
 \label{Fig:selections}
\end{figure}

The data processing starts with the beam particle reconstruction in 
$BPC_{1} \div BPC_{4}$, then the secondary tracks are looked for in 
$PC_{1} \div PC_{8}$ ; $ST_{1} \div ST_{3}$; $DT_{1} \div DT_{2}$
and events with one good positive track are selected.
The decay vertex is searched for, and a cut  is introduced on the
matching of incoming and decay track. The next step is to look for 
showers in GAMS-2000 and EGS calorimeters. The electron identification is done using 
the ratio of the energy of the shower to the momentum of the associated track. The E/p distribution is shown in 
Fig.~\ref{Fig:selections}. The particles with $0.8 < \mbox{E/p} < 1.2$ are accepted as electrons.
 The events with one charged track identified as electron and two additional 
showers in ECAL are selected for further processing. The mass spectrum of 
$\gamma \gamma$   shows a clean $\pi^{0}$ peak at  
$M_{\pi0}=134.9$ MeV with a resolution of $\sim 8.5$ MeV.
 To fight the main background from $K_{\pi 2}$ decay, the angle between the momentum of the
beam kaon $\vec{p}_{K}$ and that
of the $e \pi$-system i.e. $\vec{p}_{e} + \vec{p}_{\pi}$ is considered, see Fig.~\ref{Fig:selections}.
The background is clearly seen as a peak at zero angles. The cut is $\alpha >1.6$ mrad.
 Further selection  is done by the requirement that the event passes 2C   
 $K \rightarrow e \nu \pi^{0}$ fit. The event selection results in 
$\sim$5.25M events. The surviving background is estimated from MC to be less
than $1\%$.  
\subsection{ Analysis}
 The analysis is based on the fit of the  distribution of
the events over the Dalitz plot. The  variables 
$y=2E^*_{e}/M_{K}$ and $z=2E^*_{\pi}/M_{K}$, where $E^*_{e}$, $E^*_{\pi}$ are the
energies of the electron and $\pi^{0}$ in the kaon c.m.s are used. The
background events, as MC shows, occupy the peripheral part of the plot.  
The most general Lorentz invariant form of the matrix element for the 
decay $K^{+} \rightarrow l^{+} \nu \pi^{0}$ is  \cite{Steiner}: $ M=
\frac{-G_{F}V_{us}}{2}\bar u(p_{\nu})(1+\gamma^{5})
[((P_{K}+P_{\pi})_{\alpha}f_{+}+(P_{K}-P_{\pi})_{\alpha}f_{-})\gamma^{\alpha}-2m_{K}f_{S}-i\frac{2f_{T}}{m_{K}}
\sigma_{\alpha \beta}P^{\alpha}_{K}P^{\beta}_{\pi}]v(p_{l})$ 
 It consists of vector, scalar and tensor terms. $f_{\pm}$
are the functions of $t= (P_{K}-P_{\pi})^{2}$. In the Standard Model (SM)
the W-boson exchange leads to the pure vector term. 
The term in the vector part, proportional to $f_{-}$ is reduced(using the Dirac
equation) to a scalar form-factor, proportional to $(m_{l}/2m_{K})f_{-}$ and is negligible in the case of 
$K_{e3}$. Different parametrizations have been used  for $f_{+}(t)$. First is just a Taylor series:
$f_{+}(t)=f_{+}(0)(1+\lambda'_{+}t/m^{2}_{\pi^+} + \frac{1}{2}\lambda''_{+}t^{2}/m^{4}_{\pi^+})$. It is usually  used
to compare  with ChPT predictions. Alternative parametrization is the pole one: $f_{+}(t)=f_{+}(0)\frac{m^{2}_{V}}{m^{2}_{V}-t}$.
The last is a relatively new Dispersive parametrization \cite{dispersive}:
$f_{+}(t)=f_{+}(0) exp(\frac{t}{m^{2}_{\pi}}(\Lambda_{+}+H(t)))$. Here H(t) is a known function.\\  
The procedure for the experimental extraction of the parameters
$ \lambda_{+}$, $f_{S}$, $f_{T}$, which was developed in \cite{istraKe3} is used.
 This procedure allows avoiding  systematic errors due to the "migration" of the events over the Dalitz plot
because of the finite experimental resolution.
 The radiative corrections were taken into account by reweighting every MC event, according to \cite{Cirigliano}.
\subsection{Results and comparison with theory}
\begin{figure}[h]
  \begin{minipage}{0.50\linewidth}
    \center {\includegraphics[width=\linewidth]{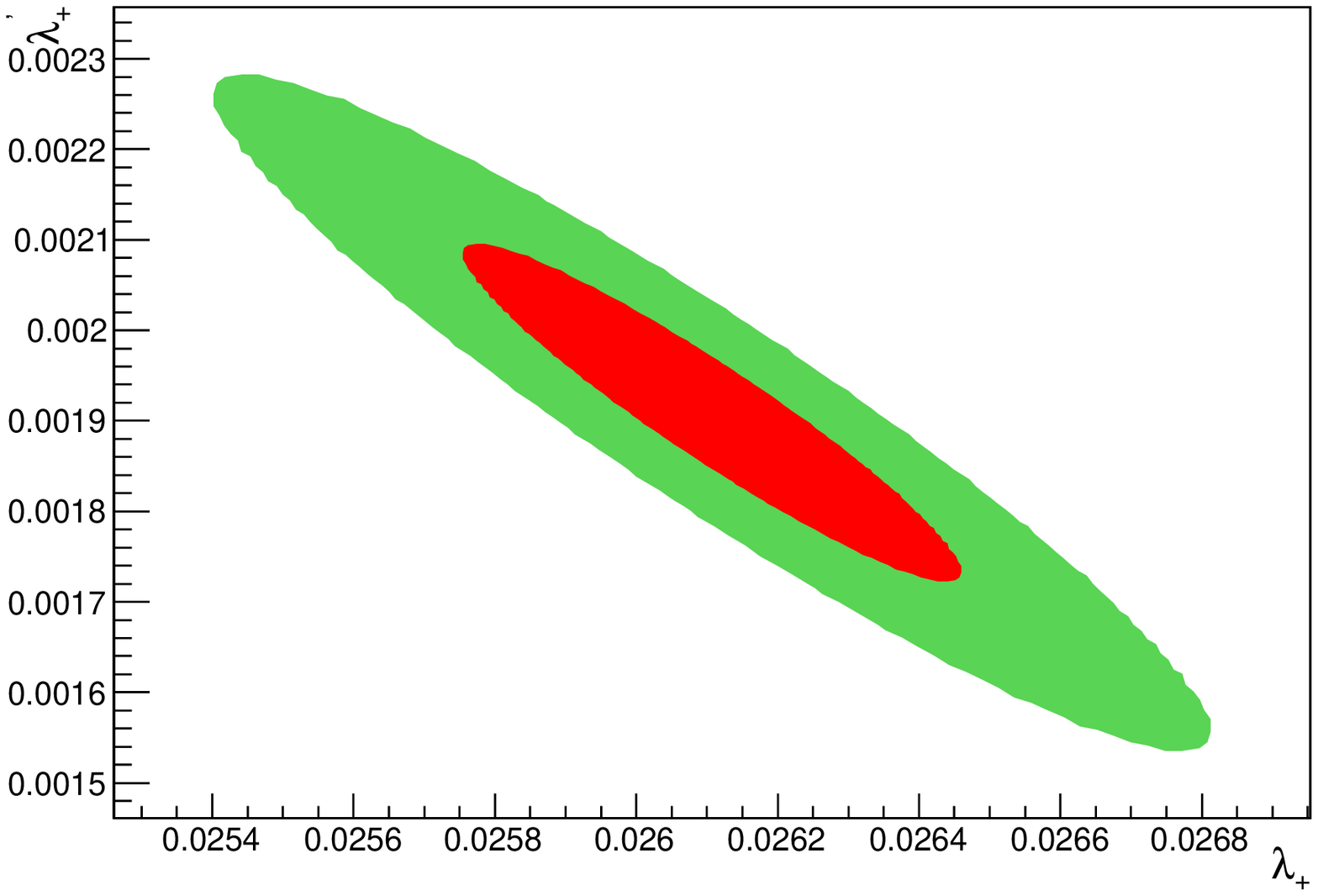}}
  \end{minipage}  
  \begin{minipage}{0.55\linewidth}
 \center {\includegraphics[width=\linewidth]{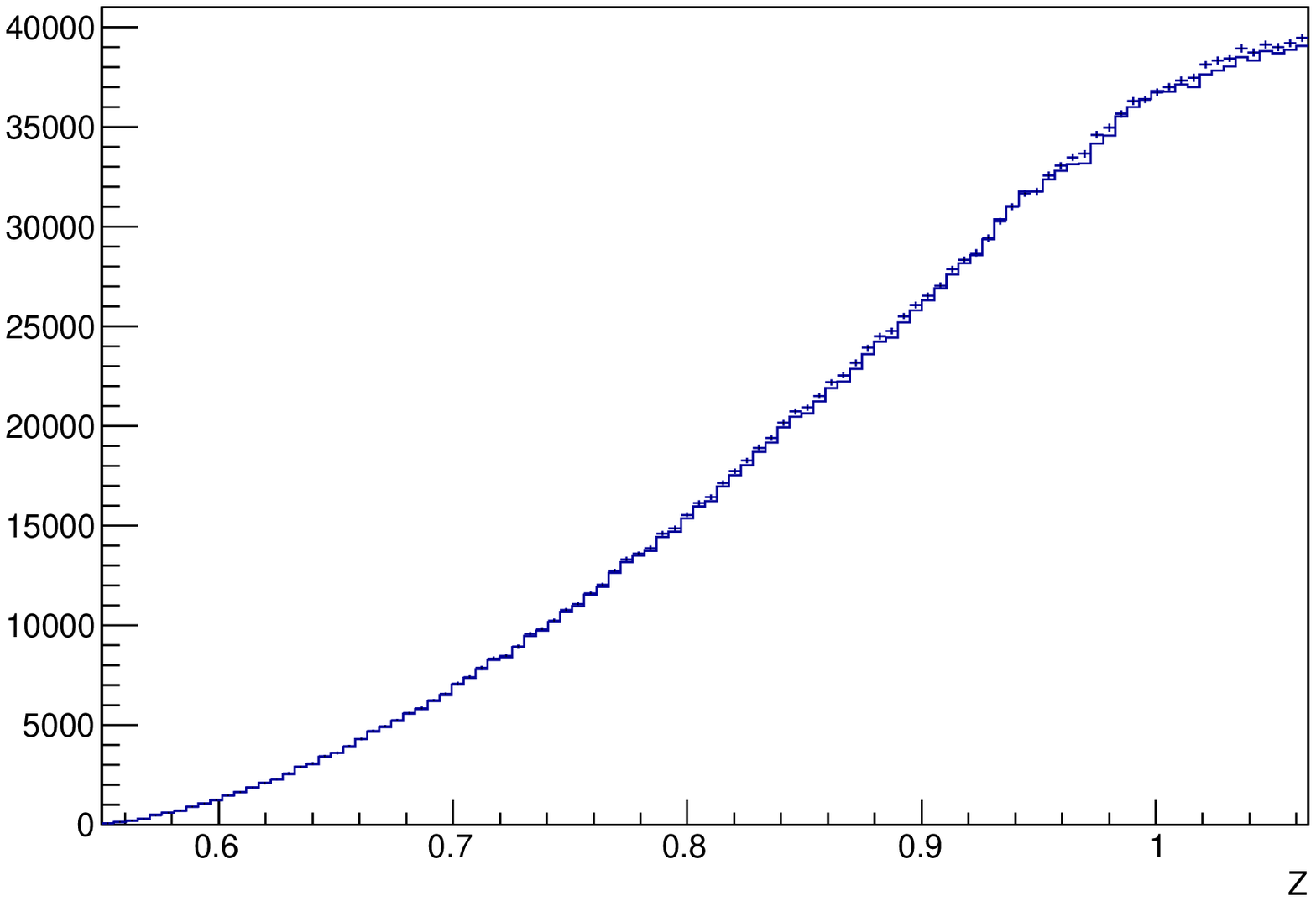}}
 \end{minipage}
 \caption{$\lambda'-\lambda''$ correlation plot (left); Projection of the Dalitz-plot on z (right) axis.
   Data is the points with errors; histogram is the fit, corresponding to the first line of the Table~\ref{tab:mainfit}.}
 \label{Fig:projections}
\end{figure}

The fit with linear parametrization of the form factor gives $\lambda_+=(2.95 \pm 0.022) \times 10^{-2}$.
 It could be compared to quite old $ChPT~ O(p^4)$ result \cite{ChPTO(p4)}:  
$\lambda^{ChPT}_+=(31.0 \pm 0.6) \times 10^{-3}$.
The results of the fits are summarized in  Table~\ref{tab:mainfit}. 
  The first line is the $``$standard$''$ fit with two parameters - linear and quadratic slopes.
The quadratic term is quite significant, there is a  strong correlation between parameters as it is seen in  Fig.~\ref{Fig:projections}.

\begin{table}
\renewcommand{\arraystretch}{1.2}
\begin{center}
\small
 \begin{tabular}{|c|c|c|c|c|c|}
 \hline
 $\lambda'_+$ $(10^{-2})$ & $m$ [GeV]& $\Lambda_+$ $(10^{-2})$ & $\lambda''_+$ $(10^{-3})$ & $f_t/f_+(0)$ $(10^{-2})$  & $f_s/f_+(0)$ $(10^{-3})$ \\ \hline
 $2.611^{+0.035}_{-0.035}$&                              &                              & $1.91^{+0.19}_{-0.18}$ &   &  \\
                                           & $0.891^{+0.003}_{-0.003}$ &                              &                  &                        &  \\
                                           &                              & $2.458^{+0.018}_{-0.018}$ &                  &                        &  \\
 $2.612^{+0.035}_{-0.035}$&                              &                              & $1.90^{+0.19}_{-0.19}$ & $-1.24^{+1.6}_{-1.3}$ & $0.13^{+3.8}_{-4.6}$  \\ 
                                           & $0.891^{+0.004}_{-0.006}$  &                            &                                      & $-1.85^{+2.4}_{-1.2}$   & $1.95^{+3.7}_{-7.4}$  \\
                                           &                              &  $2.459^{+0.019}_{-0.018}$ &                       & $-1.14^{+1.5}_{-1.3}$   & $-0.13^{+4.5}_{-3.9}$ \\ \hline
\end{tabular}
\end{center}
\caption{Results of the data fit with different possible form factors.}\label{tab:mainfit}
\end{table}

The quality of the fit is illustrated by the z  projection of the Dalitz plot, shown on 
Fig.~\ref{Fig:projections}.
 The second and third   lines of the Table~\ref{tab:mainfit}. correspond to the Pole and Dispersive fits respectively.
 Next lines represent the quadratic, Pole and Dispersive fits with additional tensor and scalar contributions. 
 It is seen, that $f_S$ and $f_T$ are not significant. 

The main contribution to systematic is coming
 from the variation of the cut on Z coordinate of the vertex and the cut on the angle $\alpha$. 
The contributions to the systematic errors from Z and $\alpha$ variations are $(0.021,\; 0.014)\cdot 10^{-2}$ and
 $(0.11,\; 0.06)\cdot 10^{-3}$ for 
$\lambda'_+$ and $\lambda''_+$ respectively.
    Finally, we get the results for the quadratic fit:   
   $\lambda'_+ = (2.611 \pm 0.035 \pm 0.028)\cdot 10^{-2}$ and 
   $\lambda''_+ = (1.91^{+0.19}_{-0.18} \pm 0.14)\cdot 10^{-3}$.

The result of the Pole fit can be compared to the PDG value for the $K^*$ mass\cite{PDG}:
 $M_{K*}=891.66 \pm 0.26$ MeV.
An interpretation of limits on $F_S$ and $F_T$ is possible in the framework of the scalar LeptoQuark(LQ)
model. Then a diagram with LQ exchange should be added to the SM diagram with W.   
Applying Fiertz transformation to the LQ matrix element we get:
 $(\bar s \mu)(\bar \nu u) = -\frac{1}{2}(\bar s u)(\bar \nu \mu)-\frac{1}{8}(\bar s \sigma_{\alpha \beta}u)(\bar \nu\sigma^{\alpha \beta}\mu) $.
The first term is the scalar, the second one - tensor. The relation between $f_S, f_T$ and the Leptoquark scale $\Lambda_{LQ}$ can be set out
(\cite{Kiselev}). As a result, a 95$\%$ lower limit for the LeptoQuark scale is   $\Lambda_{LQ} >3.5$ TeV.

\enterdate{May 29, 2017.}

\selectlanguage{russian}

\outputnames
{Ющенко О.П. и др.}
{Исследования $K_{e3}$ форм факторов в эксперименте ОКА}
{2017--2}
{авторами}
{}
{}
 \outputdata
 {01.06.2017}
 {0,62.}
 {0,67.}
 {80.}
 {4.}
 {3649.}
 {}

\end{document}